%% file: doc.tex
\pdfoutput=1
\documentclass[12pt]{article}

\usepackage{amsmath,amssymb,amsfonts}
\usepackage{graphicx}
\usepackage{cite}
\usepackage{url}
\usepackage{hyperref}
\usepackage[utf8]{inputenc}
\usepackage{algorithmic}
\usepackage{textcomp}
\usepackage{xcolor}
\usepackage{booktabs}
\usepackage{multirow}
\usepackage{subfigure}
\usepackage{float}
\usepackage{algorithm2e}
\usepackage{enumerate}
\usepackage{listings}
\usepackage{xspace}
\usepackage[left=1in,right=1in,top=1in,bottom=1in]{geometry}  
\usepackage{rotating}
\usepackage[left=1in,right=1in,top=1in,bottom=1in]{geometry}
\usepackage{pdflscape}  
\usepackage{adjustbox} 
\usepackage{tabularx}  

\usepackage{amsmath,amssymb,amsfonts}
\usepackage[utf8]{inputenc}
\usepackage{graphicx}
\usepackage{cite}

\usepackage[left=1in,right=1in,top=1in,bottom=1in]{geometry}
\usepackage{url}
\usepackage{xspace}
\usepackage{algorithmic}
\usepackage{algorithm2e}
\usepackage{booktabs}
\usepackage{multirow}
\usepackage{subfigure}
\usepackage{float}
\usepackage{enumerate}
\usepackage{listings}

\usepackage{hyperref}

\newcommand{\adae}{ADAE\xspace}
\newcommand{\slo}{SLO\xspace}

\makeatletter
\g@addto@macro{\UrlBreaks}{\UrlOrds}
\makeatother

\begin{document}

\title{Characterizing and Modeling AI-Driven Animal Ecology Studies at the Edge}

\author{
Jenna Kline \and Austin O'Quinn \and Tanya Berger-Wolf \and Christopher Stewart\\
\textit{Department of Computer Science and Engineering}\\
The Ohio State University\\
\small{\{kline.377, oquinn.18, berger-wolf.1\}@osu.edu, cstewart@cse.ohio-state.edu}
}

\date{}  

\maketitle

\begin{abstract}
\label{sect:abstract}
Platforms that run artificial intelligence (AI) pipelines on edge computing resources are transforming the fields of 
animal ecology and biodiversity, enabling novel wildlife studies in animals' natural habitats.
With emerging remote sensing hardware, e.g., camera traps and drones, and sophisticated AI models in situ, edge computing will be more significant in future AI-driven animal ecology (ADAE) studies.
However, the study's objectives, the species of interest, its behaviors, range, and habitat, and camera placement affect the demand for edge resources at runtime.
If edge resources are under-provisioned, studies can miss opportunities to adapt the settings of camera traps and drones to improve the quality and relevance of captured data.
This paper presents salient features of ADAE studies that can be used to model latency, throughput objectives, and provision edge resources.
Drawing from studies that span over fifty animal species, four geographic locations, and multiple remote sensing methods, we characterized common patterns in ADAE studies, revealing increasingly complex workflows involving various computer vision tasks with strict service level objectives (SLO).
ADAE workflow demands will soon exceed individual edge devices' compute and memory resources, requiring multiple networked edge devices to meet performance demands.
We developed a framework to scale traces from prior studies and replay them offline on representative edge platforms, allowing us to capture throughput and latency data across edge configurations.  
We used the data to calibrate queuing and machine learning models that predict performance on unseen edge configurations, achieving errors as low as 19\%.

\end{abstract}


\section{Introduction}
\label{sect:introduction}
\begin{figure}[t]
    \centering
    \includegraphics[width=0.8\linewidth]{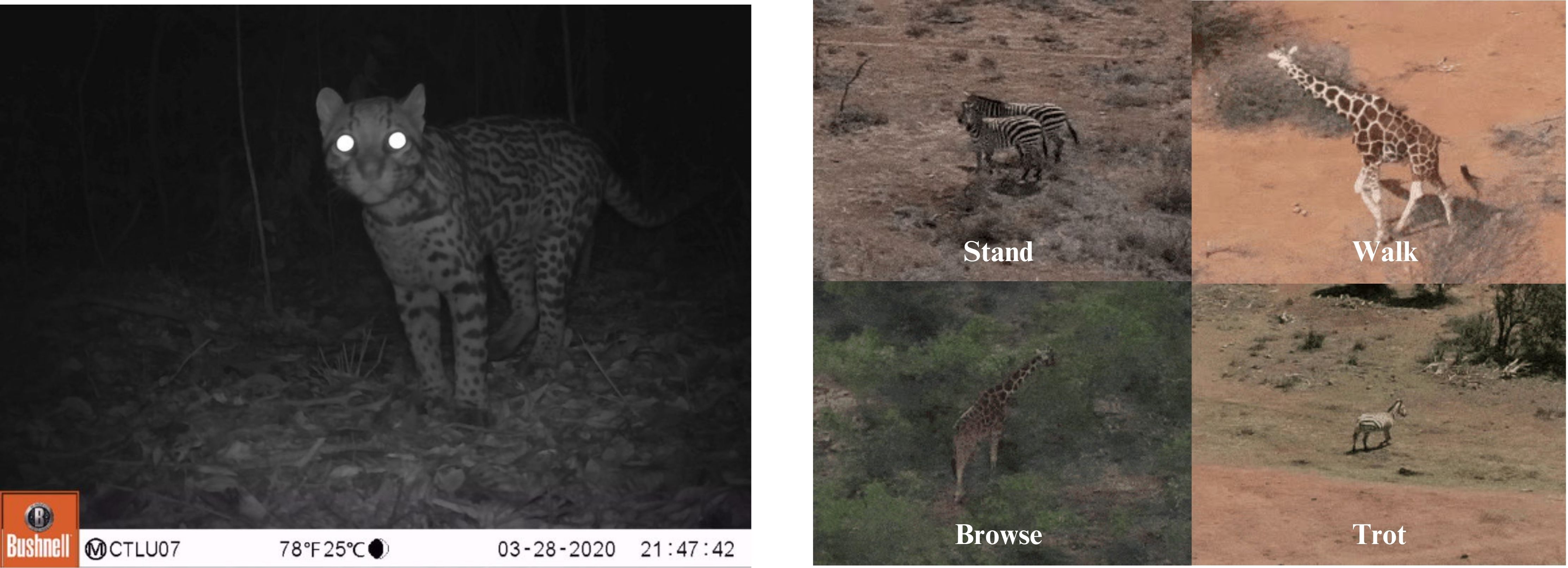}
    \caption{Data captured from animal ecology studies in the field: 1) A camera trap captures large cats' species range and nightly activities, adapted from ~\cite{Vélez_McShea_2023}. 2) Drones capture animal behavior and poses, adapted from ~\cite{kholiavchenko2024kabr}. }
    \label{fig:photos}
\end{figure}

Camera traps and drones can automatically capture visual data on animals, their morphology and behaviors, and biodiversity within an ecosystem, transforming the fields of animal ecology and biodiversity (Figure~\ref{fig:photos}).
It is now common for field-based animal ecological studies to use more than 70 camera traps~\cite{Burton_Neilson_Moreira_Ladle_Steenweg_Fisher_Bayne_Boutin_2015}. \cite{Vélez_McShea_2023}.
Between 2015 and 2020, at least 19 academic studies were driven by aerial drone imagery~\cite{Corcoran_Winsen_Sudholz_Hamilton_2021}.
Drones are especially promising for animal behavior studies that require tracking wildlife over vast, remote landscapes~\cite{Schad_Fischer_2023, Kholiavchenko_Kline_2024, Duporge_Kholiavchenko_Harel_Wolf_Rubenstein_Crofoot_Berger-Wolf_Lee_Barreau_Kline_et_2024, Ozogány_Kerekes_Fülöp_Barta_Nagy_2023, Koger_Deshpande_Kerby_Graving_Costelloe_Couzin}.
Computer vision and machine learning approaches have sped up post hoc processing for visual data collected in the field~\cite{Tuia_Kellenberger_Beery_Costelloe_Zuffi_Risse_Mathis_Mathis_Van_Langevelde_Burghardt_et, Norouzzadeh_Nguyen_Kosmala_Swanson_Palmer_Packer_Clune_2018, Brookes_Mirmehdi_Kühl_Burghardt_2023, kholiavchenko2024kabr, Young_Rode-Margono_Amin_2018,MegaDetector-WILDLABS}.
However, as camera traps and drones flood ecologists with data, it is challenging to curate, process, and manage the data to discover ecological insights in a timely fashion~\cite{Fergus_Chalmers_Longmore_Wich_Warmenhove_Swart_Ngongwane_Burger_Ledgard_Meijaard_2023, Zualkernan_Dhou_Judas_Sajun_Gomez_Hussain_2022,Whytock_Suijten_van}.
%
Further, AI pipelines that infer complex ecological traits require
images with prescribed pixel resolution, angles, and timing: factors 
related to data quality that is determined at runtime.
Images with low resolution or occlusions require 
expert analysis to decipher insights or must be discarded altogether.

Edge AI, the application of AI pipelines on edge computing systems~\cite{Singh_Gill_2023}, can enable AI-driven animal ecology (ADAE) studies.
ADAE studies control remote sensing systems at runtime, filtering images, adjusting angles, and changing camera or drone positions to improve data quality through adaptive sampling~\cite{Whytock_Suijten_van, Zualkernan_Dhou_Judas_Sajun_Gomez_Hussain_2022, Dertien_Negi_Dinerstein_Krishnamurthy_Negi_Gopal_Gulick_Pathak_Kapoor_Yadav_2023, Chalmers_Fergus_Wich_2023, Fergus_Chalmers_Longmore_Wich_Warmenhove_Swart_Ngongwane_Burger_Ledgard_Meijaard_2023, Henrys_Mondain-Monval_Jarvis_2024}.
Ecologists are beginning to use edge AI platforms to conduct ADAE studies using networks of smart camera traps \cite{Tulasi_Granados_Gunawardane_Kashyap_McDonald_Thulasidasan_2023, TheSentinel, Animl}.
Drones are innately adaptive if they are piloted well.  
Edge AI can reduce the burden on pilots, allowing ADAE studies to employ multiple drones, capture data from vast areas, and improve data quality~\cite{Kline_Kholiavchenko_Brookes_Berger-Wolf_Stewart, Luo_Zhang_Shao_Zhao_Wang_Zhang_Liu_Li_Liu_Wang_et_2024, boubin2022, bala2023democratizing}.

Animal ecology studies collect data from predefined locations by placing camera traps or flying predefined missions. 
Our study provides a critical insight: {\em adaptive data collection enabled by edge computing can improve study efficacy.} 
We present the first characterization of ADAE studies.
ADAE study workflows employ image analytics at the edge, composing inter-dependent, inference pipelines from complex AI computer vision models~\cite{jang2021microservice,yi2017lavea}.  
These workflows must be executed under strict SLOs to support runtime adaptations. 
In this paper, our contribution is a characterization of ADAE studies, their definitive features, workload demands, and the factors affecting their performance.

The remainder of the paper is organized as follows.
We characterize ADAE workloads in Section~\ref{characterizing} and describe our methodology to model and scale ADAE workloads in Section~\ref{modelling}.
Section~\ref{sect:experiments} describes our framework for experimenting on representative hardware and analyzes factors affecting SLO attainment for prior studies.
Section~\ref{sect:results} examines performance modeling for ADAE studies.
Section~\ref{sect:relatedwork} reviews related works.
Section~\ref{sect:conclusion} summarizes our findings and future work.

\section{Characterization of ADAE Studies}
\label{characterizing}

We analyze datasets from prior AI-driven animal ecology (ADAE) studies to characterize their workloads \cite{kholiavchenko2024kabr, Vélez_McShea_2023}.
However, instead of searching for ecological insights, we examine when the data was collected and what software components were triggered using timestamps provided by camera traps and drones.
To our knowledge, this work is a first attempt to apply ADAE study traces to profile the characteristic workload demands from an edge perspective.
Our analysis of the frequency and timing of timestamps reveals that image capture and subsequent computational triggers occur in bursts.
Further, approaches to expand a study's geographic footprint affect the magnitude of bursts.  
Timestamp analysis also revealed the latency window for edge computing systems to make runtime adaptations to improve data quality. 
We adapted service-level-objectives (SLO) to characterize 
ADAE study demands a widely used paradigm in cloud computing.

\subsection{ADAE workflow: design, execution, and results}
\label{sect:workflow}

\begin{figure*}
    \centering
    \includegraphics[width=1\linewidth]{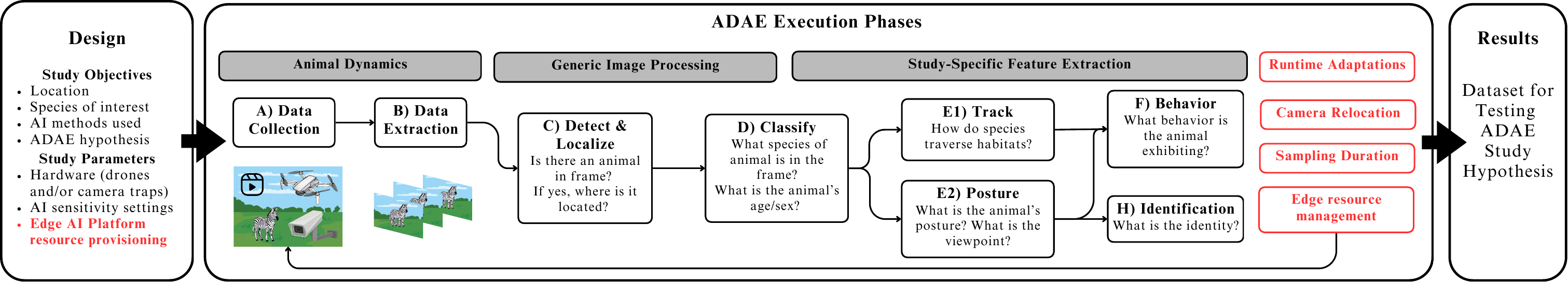}
    \caption{Canonical workflow for field AI-driven animal ecology (ADAE) studies.}
    \label{fig:cv_pipeline}
\end{figure*}

We illustrate the canonical workflows of ADAE studies in Figure~\ref{fig:cv_pipeline}. 
The ADAE workflow comprises three phases: design, execution, and results.
The design phase consists of establishing the study objective and study parameters.
The ADAE study objectives include the location, species of interest, AI methods used, and ADAE hypothesis.
The study parameters include the remote sensing hardware used, such as drones or camera traps, the AI sensitivity settings, and the edge resource provisioning strategy.
The ADAE execution workflow includes four subphases: (1) animal dynamics, (2) generic image processing, (3) study-specific feature extraction, and (4) runtime adaptations.
The final phase produces the results, where the dataset has been collected and is ready for analysis.

The first subphase of ADAE study execution is \textit{animal dynamics}.
This includes collecting imagery data with drones or camera traps of the animals of interest and extracting the collected frames for analysis.
In this phase, the data arrival rate is dictated by the behavior of the species of interest and its interactions with the camera trap and drone hardware.
The second subphase is \textit{generic image processing}.
This includes detection and localization computer vision tasks to answer the following questions: Is there an animal in this frame? If so, where is the animal located in the frame?
The classification computer vision task may be viewed as a component of the generic image processing tasks if required for a downstream task, like individual identification. Classification may also be considered a study-specific feature extraction if used to complete a biodiversity ADAE study.
Commonly used computer vision models for detection, localization, and classification tasks for ADAE include YOLO~\cite{yolov8_ultralytics} and PyTorch Wildlife~\cite{hernandez2024pytorchwildlifecollaborativedeeplearning}.

The third subphase is \textit{study-specific feature extraction}. This includes computer vision tasks to infer information, including the animal's tracks, posture, behavior, and individual identification~\cite{WildMe}.
These study-specific feature extraction tasks inform the fourth subphase, \textit{runtime adaptations}, which may include camera relocation, adjusting the sampling duration, and updating the edge resource management to respond to the workload demands.
The runtime adaptation instructions are returned to the data collection module, and the ADAE study continues execution until sufficient data has been collected.

Unlike traditional field ecological studies, ADAE workloads require computational resources provisioned at the edge.
Like traditional studies, ADAE studies can fail because the data is insufficient to support or reject the hypothesis.  
However, ADAE studies only succeed if the edge platform can make runtime adaptations quickly enough to capture high-quality study-appropriate data.
We do not claim that our study is representative of all ADAE studies, but the observed characteristics are well-motivated and will likely generalize to future studies.

\input{figures/table.tex}

\subsection{Required SLOs for runtime adaptations}
\label{sect:slo}
The latency requirement for the ADAE study is dictated by the service-level-objective (SLOs) of the computer vision pipeline used to gather the data and inform runtime adaptations.
The computer vision pipeline for ADAE workloads is illustrated in Figure \ref{fig:cv_pipeline}.
SLOs for specific ADAE studies are detailed in Table \ref{table:ae_slos}. 
An additional component of the SLO is the rate of requests met or the percentage of inference requests that must be met.
The required rate of requests met varies depending on the computer vision tasks and study parameters.
Depending on the edge hardware available and the study design, some computer vision tasks may have strict SLO. At the same time, other pipeline components may be offloaded to the cloud for post-hoc analysis.
The average number of frames, from photos or video, that the computer vision pipeline must process per second determines the SLO requirements.
Depending on the study design, the pipeline may include one or more computer vision models to accomplish different tasks.
For computer vision pipelines that navigate autonomous drones for ADAE, the SLO will be stricter than studies using camera traps~\cite{boubin2019autonomic,boubin2022,bala2023democratizing}.
For camera trap studies, the SLO is set to handle the real-time data analysis and processing needed to inform runtime adaptations, such as the frequency and sampling duration.

A secondary benefit of the ADAE approach is that it can enable near-real-time ecological insights instead of solely relying on post hoc processing techniques.
Here, near real-time means completing a request to process an image in minutes to hours (versus days to months) from when the image was first collected.
If no edge processing is used, the large volumes of imagery captured must all be processed offline after the study is concluded, which may take months or years to analyze.
Moving the detection and localization computer vision tasks to the edge hardware reduces the amount of data that must be analyzed offline.
For example, the Orinoquía camera trap dataset contains 20\% blank imagery \cite{Vélez_McShea_2023}.
Current state-of-the-art for near real-time camera trap image processing is an average of 7.35 minutes per image for a network where each camera produced 17 images per day on average~\cite{Whytock_Suijten_van}. 

\subsection{Representative ADAE studies}
\label{sect:studies}
We describe seven representative ADAE studies in Table \ref{table:ae_slos}.
We use the hardware, species of interest, data type, computer vision tasks, and desired outputs to define the SLO requirements for the ADAE studies.
ADAE 1 uses a fixed-wing drone to survey bison to count the number of calves present in the herd \cite{Corcoran_Winsen_Sudholz_Hamilton_2021}.
Counting the number of young is an important data point to quantify the success of conservation efforts to repopulate this species in the American plains. 
We assume the fixed-wing drone uses the default settings for a survey mission to generate an orthomosiac image of the herd: nadir-view, 60 m altitude, 1.3 cm/pixel resolution, with a 75\% front overlap and 70\% side overlap \cite{dronedeploy}. 
This flight plan generates approximately one image every 0.2 seconds, 
however, due to the overlap in images, it is sufficient to analyze 30\% of the images received and still be confident that the bison are in view of the drone.
ADAE 2 also uses a single fixed-wing drone to detect the presence of endangered wildlife in a conservation area.
This approach is similar to the SPOT Poachers in Action study~\cite{iot_Fang_Hamilton_Kar_Dmello_Choi_Hannaford_Iyer_Joppa_Tambe_etal._2018}, which reported an average latency rate of approximately 1 second per frame with a GPU.
Endangered animals with low population levels may be rarely spotted, therefore, the minimum requests met for such studies is high to ensure the frames containing rare or endangered species are not missed.

ADAE 3 uses a single quadcopter drone to count the number of giraffes present in different habitats in Kenya by detecting, localizing, and classifying the animals.
The imagery collected by the drone is also used to classify the habitat as open or closed, categorized by the amount of vegetation present. 
A quadcopter drone was selected for this study because, unlike fixed-wing drones, it can more easily navigate around occlusions from vegetation in closed habitats.
Group-living animals may be autonomously tracked with drones using a detection and localization model, such as YOLO, integrated into the control software~\cite{Kline_Stewart_Berger-Wolf_Ramirez_Stevens_Babu_Banerji_Sheets_Balasubramaniam_Campolongo_2023}.
This autonomous herd-tracking navigation model requires a 1-second per frame latency and a tolerance of 80\%. 
The frame rate may be adjusted depending on the average speed of the species of interest. 
If the SLO is violated, the drone may lose sight of the animals, forcing the data collection mission to end prematurely.
ADAE 4 scales ADAE 3 by implementing the herd-tracking navigation pipeline with a swarm of multiple quadcopters.
The aim of ADAE 4 is to collect videos of zebra herds to study their behavioral differences by the time of day,  similar to the methodology used in the KABR study \cite{kholiavchenko2024kabr}.
As this study collects behavior videos of group-living animals instead of photos of a single species, it requires a longer sampling time compared to ADAE 3 but maintains the same SLO.

ADAE 5 represents a single smart-camera trap study that collects video behavior data of a pack of African Wild Dogs at a wildlife conservation center.
This study uses a motion-activated camera to trigger video recording if an animal is detected in view. The camera tracks the animal until it is out of sight.
This methodology for collecting behavior videos with motion-activated smart camera traps has been successfully used to collect large-scale ape behavior datasets \cite{Brookes_Mirmehdi_Kühl_Burghardt_2023, VAIB}.
For ADAE 5, the recording duration is the essential runtime adaptation, which depends on the accuracy of the tracking step (E1 shown in Figure \ref{fig:cv_pipeline}).
For this study, African Wild Dogs are the only species in the enclosure, so species classification is unnecessary.

ADAE 6 uses a single, smart camera trap to collect photos to estimate species distribution and population estimates for a biodiversity study.
It is assumed an average of 20 images are collected each hour, so the SLO for completing the CV pipeline is 3 minutes per image, or 600 seconds per frame, to prevent a queue from being formed.
ADAE 7 independently scales the study from ADAE 6 by adding additional smart camera traps distributed geographically to estimate species distributions and populations over a wider area, similar to the Orinoquía Camera Trap study~\cite{Vélez_McShea_2023}, which we obtained from LILA BC repository online \cite{Lila}.

ADAE 8 scales the study from ADAE 6.
Instead of placing camera traps distributed geographically, it places additional camera traps in the same spot, which produces correlated scaling.
This correlated scaling approach is better suited for behavioral studies (F from Figure \ref{fig:cv_pipeline}) and studies requiring individual identification (H from Figure \ref{fig:cv_pipeline}) using a tool like WildMe \cite{WildMe}.
AI computer vision models to classify behavior and identify individual animals benefit from having access to views of the animal(s) from multiple angles, which requires a correlated scaling approach.

\subsection{ADAE workloads in Edge AI research}
\label{sect:useinedgeai}
Our group and others collected the ADAE traces profiled in this work.
However, all ADAE traces profiled contained two essential components. One, imagery data in the form of videos or photos. Two, timestamped arrival rates for the imagery data associated with request arrivals for the ADAE computer vision tasks.
Arrival rates for computer vision tasks depend on the specific ADAE study objectives, as described in Table~\ref{table:ae_slos}.
An essential contribution of our effort is discovering commonalities that enabled rigorous analysis.
Edge systems researchers can leverage the SLOs described in this section to explore new distributed computing techniques designed for use in field ADAE studies.
By analyzing the workload patterns revealed by timestamp data, researchers can focus on optimizing SLOs for edge computing systems.
This could include developing techniques to predict and manage latency for ADAE-specific computer vision tasks and ensuring runtime adaptations occur within the required time frame for improved data quality.

\section{Modeling ADAE Workloads}
\label{modelling}
Our workload characterization of ADAE studies in Section \ref{characterizing} suggests that their computational demand will exceed the capacity for individual edge devices deployed in remote settings. 
Meeting latency and throughput goals will require assessing edge configurations before deployment and predicting their performance before resources are provisioned.
Evaluating proposed configurations in situ is challenging due to ethical, logistical, and resource constraint considerations; thus, we present a framework to enable offline evaluation. 

We model ADAE workloads using the characteristic request arrival rates generated by these studies described in Section~\ref{characterizing}.
We describe study features that affect ADAE workload burstiness: ecological factors, camera placement and scaling, and hardware and AI model considerations.
We describe our methodology to characterize and quantify these bursty workloads as a time-varying Poisson process.
We also provide code to profile and scale real-world ADAE studies along with worked examples here: 
\href{https://github.com/jennamk14/adae\_model}{https://github.com/jennamk14/adae\_model}.

\subsection{Factors driving burstiness}
\label{sect:burstyfactors}
Computer vision model workloads often exhibit burstiness, with periods of high activity followed by low or no activity intervals, depending on the ADAE study parameters \cite{Weinstein_2018}.
These study parameters include the type and location of the sensors, the AI models used, and the habitat and behavior of the species of interest.
Bursty workloads describe those in which request arrival times are unpredictable, but there is also a high degree of covariance between requests.
Previous studies demonstrate that autonomous navigation models that track and monitor animals using drones produce bursty workloads \cite{kholiavchenko2024kabr}. 
Bursts of high arrival rates increase the queuing times and processing delays, potentially violating \slo.
When determining \slo for ADAE studies, we aim to minimize queuing delays to meet the latency requirements. 
Thus, the bursty nature of these workloads must be considered when designing systems to meet these requirements.

\subsubsection{Ecological factors}
\label{sect:ecologicalfactors}
Various factors can influence burstiness, including species-specific activity patterns, seasonal variations, and inter-species interactions. 
%
%
Species that share the same space may actively interact with each other (e.g., predation or resource competition), neutrally coexist (e.g., mixed-species groups of ungulates in the Serengeti \cite{Sinclair_1985}), or actively avoid each other (e.g., tigers and leopards \cite{Karanth_Sunquist_2000}).
Active species interactions are rare and require overlapping workloads covering the potential interaction occurrence area. 
Neutral coexistence leads to overlapping detection and classification model workloads.
Finally, active avoidance leads to mostly non-overlapping workloads.

Species-specific activity patterns influence burstiness.
By definition, diurnal species are active during the day, and crepuscular species are active during dawn and dusk, generating camera trap captures during different times of day.
%
%
Thus, quantified using the coefficient of variation metric, diurnal patterns would considered bursty. 
However, animal ecology patterns are less pronounced and predictable than diurnal patterns in cloud or e-commerce systems.

\subsubsection{Camera placement and scaling strategies}
\label{sect:scaling}
Placing camera traps and drones significantly shapes the workload dynamics and determines the appropriate scaling strategies. 
Two common scaling approaches are 1) Independent scaling, which distributes more cameras over a large area, and 2) Correlated scaling, which increases the camera density in specific locations.
Independent scaling is typically used to study wide-ranging species, such as wolves or migratory ungulates, to understand their landscape-level movements and habitat preferences \cite{Mech_Boitani_2010}, or studying the spatial distribution of sympatric species, such as jaguars and pumas \cite{Harmsen_Foster_Silver_Ostro_Doncaster_2009}.
By distributing the camera traps or drones over a large area, independent scaling is suitable for studies focusing on species distribution, habitat use, or landscape-level interactions \cite{Rovero_Zimmermann_2016}. 
Independently scaling the hardware increases the spatial coverage and captures a broader range of animal activities, reducing the burstiness of the workload. 
However, it may lead to increased workload overlap as different species' territories or movement patterns are more likely to be captured simultaneously.

Correlated scaling by increasing the camera density in specific locations is appropriate for studies focusing on fine-scale animal behavior using drones or camera traps \cite{Ozogány_Kerekes_Fülöp_Barta_Nagy_2023, Brookes_Mirmehdi_Kühl_Burghardt_2023}, social behavior and group dynamics of species like zebras~\cite{kholiavchenko2024kabr}, chimpanzees, or African elephants, which require detailed observations at specific sites~\cite{Kühl_Kalan__2016}. Inter-species interactions, or monitoring hotspots of activity \cite{Kays_Kranstauber_Jansen_Carbone_Rowcliffe_Fountain_Tilak_2009}, such as water holes or mineral licks where multiple species congregate, allowing for the study of inter-species interactions and temporal partitioning of resources \cite{Jamadagni_2012}.
This approach increases burst intensity during events, as multiple cameras capture the same activity from different angles or close succession. 

\subsubsection{Hardware and AI models} 
The type of hardware, e.g., smart camera traps or drones, and the AI computer vision models used to analyze the data impact the study's workflow.
Camera traps and drones may continuously record data, generating a constant data stream for analysis.
Or, more commonly for camera trap studies, use a motion or heat-activated sensor to capture photographs or videos only when an animal is present \cite{Hamann24cvpr}.
Fixed-wing drones are typically deployed to survey extensive, remote areas and capture photographs, which are analyzed to detect and classify the animals \cite{Hua_2022, SoftwarePilot}.
The workflow generated by fixed-wing drone missions depends on the frequency at which the drone captures animals, which is impacted by the habitat and species of interest.
Quadcopter drones are smaller and more agile, allowing them to follow groups of animals and quickly navigate to capture a variety of angles, which are particularly effective for behavior studies \cite{Kline_Kholiavchenko_Brookes_Berger-Wolf_Stewart, kholiavchenko2024kabr}.
The autonomous navigation models used to pilot these drones often exhibit bursty characteristics \cite{kline_acsos}.

The workflow of the ADAE study is also impacted by the computer vision tasks performed on the edge to enable the required runtime adaptations, as shown in Figure \ref{fig:cv_pipeline}.
Adjusting recording duration in camera trap studies can be dynamically adapted based on the species or behavior detected.
For drone studies, runtime adaptations include the navigation decisions based on the detected species or behavior~\cite{Kline_Kholiavchenko_Brookes_Berger-Wolf_Stewart, McNutt_Zhang_Carey-Douglas_Vollrath_Pope_Brickson, kholiavchenko2024kabr}. 

\subsection{Methodology for modeling workloads}
\label{subsec:burstiness}
We model workloads as time-varying Poisson processes, with rate changes identified at key inflection points. 
This allows the scaling of traces while maintaining the burstiness characteristics. 
This method preserves the realism of animal ecology workloads while enabling the testing of different hardware and configurations through workload generation based on real-world traces.

\begin{table}[t]
    \centering
    \begin{tabular}{lll}
    \toprule
    \textbf{Number of Cameras} & \textbf{ADAE Study (Table \ref{table:ae_slos})} & \textbf{CoV}  \\
    \midrule 
    Single smart camera trap & 5 & 1.59 \\
    Smart camera trap network & 7 & 7.33\\
    Single quadcopter  & 3 & 3.01  \\
    \bottomrule \\ 
    \end{tabular}
    \caption{Quantifying Burstiness of ADAE Studies}
    \label{tab:burstinessmeasures}
\end{table}

\begin{figure}
    \centering
    \includegraphics[width=1\linewidth]{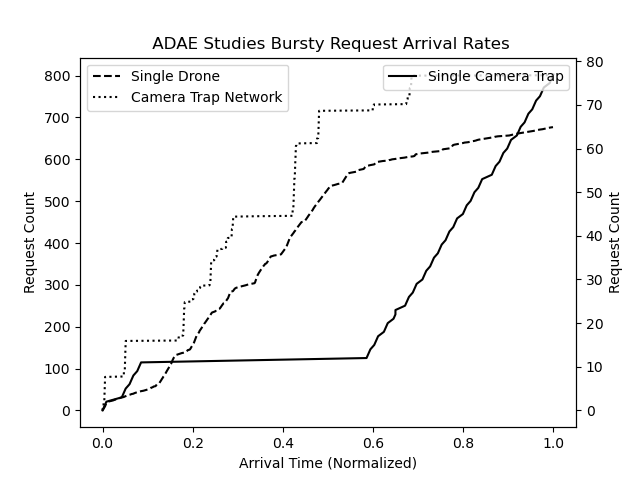}
    \caption{Modeling the burstiness of ADAE studies. The request count for the Single Drone and Camera Trap network traces are shown on the left y-axis, and the request count for a Single Camera trap is shown on the right y-axis. Arrivals are normalized with respect to time- the single camera trap arrival data has a 24-hour duration, while the drone and camera trap arrival times have a 2-hour duration. }
    \label{fig:bursty_workload_plot}
\end{figure}

\subsubsection{Quantifying burstiness}
We characterize the burstiness of ADAE workloads by the time each burst arrives, $t \in T$, burst duration $\mu$, and arrival rate within a burst $\lambda$.
We quantify the burstiness of ADAE using the coefficient of variation, a metric commonly used to characterize bursty arrival rates \cite{adegboyega_2017}, with results shown in Table \ref{tab:burstinessmeasures}.
The burstiness of three representative ADAE studies are visualized in Figure \ref{fig:bursty_workload_plot}, where portions with a relatively larger gradient represent a burst.
A single, smart camera trap (ADAE 5) exhibits the least burstiness, as there are only three change points where the arrival rates change dramatically, reflected in the CoV score of 1.59.
ADAE 3 with a single quadcopter exhibits comparatively more bursts than ADAE 5, where the gradient increases more rapidly with a CoV of 3.01.
A network of smart camera traps, ADAE 7, exhibits the highest level of burstiness, visualized in Figure~\ref{fig:bursty_workload_plot} as steep gradients where the arrival rate increases rapidly, which is reflected in its CoV of 7.33. 
In practice, these workloads may scale, for example, from 40\% to 80\% utility of a single node, due to independent or correlate scaling, as discussed in Section \ref{sect:scaling}.

\subsubsection{Modelling and scaling bursty workloads}
We model the bursty workloads as Poisson processes with rate variations at given change points
%
to preserve the characteristic burstiness in the arrival rates. 
To model the traces as a time-varying Poisson process, we identified the inflection points of the arrival rate gradient, denoted as change points. These change points define trace segments and the average arrival rate $\lambda$  was calculated for each segment duration. To scale the traces, we multiplied all $\lambda$ in the trace by a factor to generate the desired average utilization. 
This approach allows to capture the expected scaling from different animal ecology studies. For example, better cameras with a higher frame rate, larger models, or slower hardware can be modeled while maintaining the shape of our arrival curve. 
The scaling approach focuses on the relative parametrization, normalizing among realistic animal ecology study characteristics while testing different configurations.

\subsubsection{Workload generation}
The inputs of our bursty workload generator is the total time of the simulation $T$, and rate $R_{t}$, where $t$ is the change point, and \( t \in T \). 
The process to generate bursty arrival times from real-world traces is illustrated in Algorithm \ref{alg:wl}. 
For each unique change point value $t$, we calculate the duration of the $R_{t}$, corresponding to $t$. 
The arrivals are generated by randomly sampling the Poisson distribution of $R_{t}$ multiplied by the duration. 
Next, the arrivals are uniformly sampled for the timestamps for the duration of the rate $R_{t}$. 
These arrivals are sorted and added to the array $t_{a}$. Finally, the arrival times for the last change point's interval $t$ are generated similarly.

\begin{algorithm}[H]
    \KwData{$T$ (total time of the simulation), $R_{t}$ (rate), where $t$ are the change points, and $t \in T$.}
    \KwResult{List of bursty arrival times.}
    \BlankLine
    \caption{Bursty arrival times modeled as Poisson process with rate variation}
    $t_{s} \gets [] $ *\tcp{start time of current duration}
    $t_{a} \gets []$ *\tcp{arrival times}

    \For{$i$ \textbf{in} $t$}{
    $rate = R_{i}$ \\
    $duration = i - t_{s}$ \\ 
    $arrivals \overset{{\scriptscriptstyle \operatorname{R}}}{\leftarrow}  Poisson(rate*duration)$ \\
    $arrivals \overset{{\scriptscriptstyle \operatorname{R}}}{\leftarrow}  U (t_{s}, t_{s} + duration, arrivals) $ \\
    $t_{a} \gets $ \textbf{sort} $arrivals$
    }
    *\tcp{handle last interval}
    $rate = R_{-1}$ *\tcp{last rate in list}
    $duration = T - t_{s}$ \\
    $arrivals \overset{{\scriptscriptstyle \operatorname{R}}}{\leftarrow}  Poisson(rate*duration)$ \\
    $arrivals \overset{{\scriptscriptstyle \operatorname{R}}}{\leftarrow}  U (t_{c}, t_{c} + duration, arrivals) $ \\
    $t_{a} \gets $ \textbf{sort} $arrivals$ \\
    \textbf{return} $arrival\_times$ 
\label{alg:wl}
\end{algorithm}

\begin{figure*}
    \centering
    \includegraphics[scale=0.65]{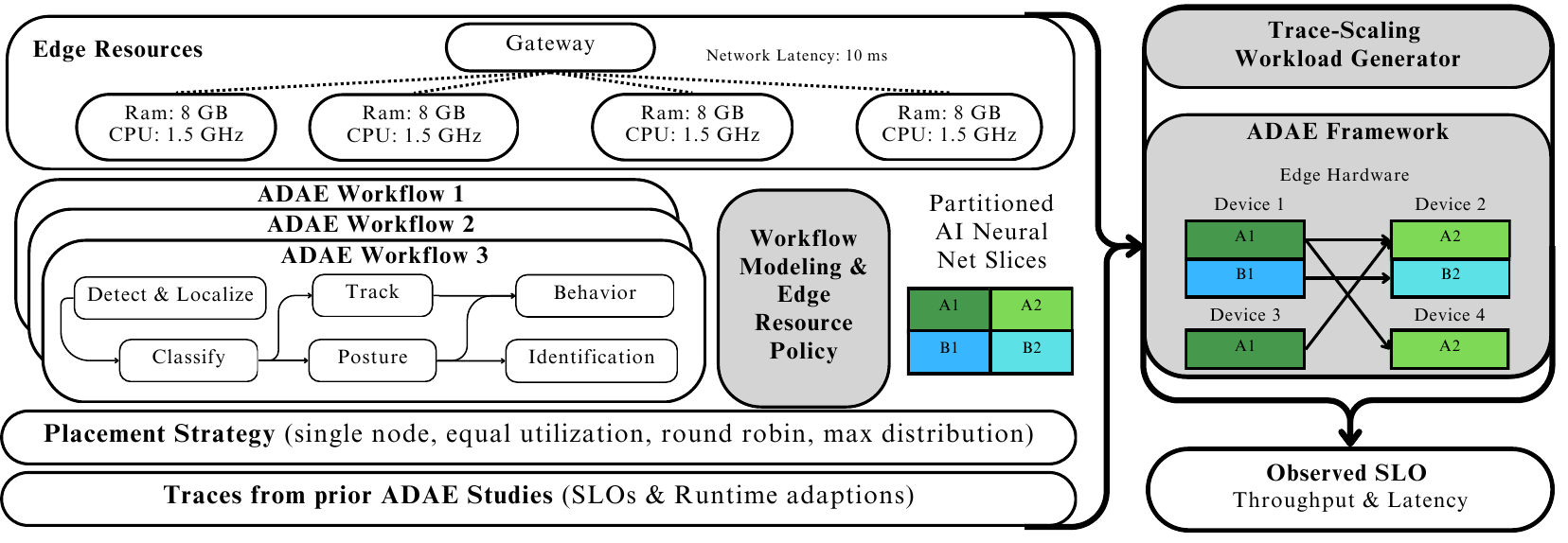}
    \caption{Framework for characterizing latency and throughput on edge AI platforms for \adae studies}
    \label{fig:draw-camtraparchitecture}
\end{figure*}

\section{Framework for Scaling and Replaying Traces from \adae Studies}
\label{sect:experiments}

Replaying and scaling \adae studies in situ presents ethical
and logistical issues.
Ethically, deploying camera traps and drones in the field 
can disturb natural habitats, discomfort animals, and provide
pathways for poachers to victimize protected species.
One-off, long-term deployments yielding valuable ecological 
insights can address these concerns, but throughput and 
latency tests do not justify the ethical risks. 
Logistically, \adae studies are conducted in remote areas
away from research labs and electrical power infrastructure.
Replaying studies in situ to test edge configurations imposes
a significant logistical burden. 
However, for \adae studies, under-provisioned edge resources hamper
runtime adaptations, leading to inconclusive study outcomes. 
It is critical to test edge configurations before studies 
begin under realistic conditions.
As discussed in Section~\ref{characterizing}, edge configurations
for future studies will likely need to support (1) bursty traffic, 
(2) multiple, co-located workflows seeking different outcomes 
(e.g., population counts and behavior profiles), and 
(3) large and complex computer vision models.

Our framework considers edge environments comprising multiple networked nodes that share computational resources to meet aggregate demand.
These edge resources can use distributed inference, i.e., partitioning workflows and placing partitions on specific nodes for execution,  to improve performance~\cite{hui2021characterizing,jang2021microservice,hsu2019couper}.  
Given edge resources and a trace from a prior \adae study, our framework can set up and test the workflow offline on representative hardware. In addition, our framework can partition and automatically distribute workflows, scale traffic from camera traps and drones, support co-located studies, and test various network latency settings.
We also developed a predictive model to forecast workload performance and service-level-objective (\slo) attainment for different edge configurations and placement strategies. 
The model accurately predicts how workloads will perform. 
These predictive capabilities enable informed decisions on resource provisioning and system deployment for \adae studies.

Using representative edge hardware, we developed a framework to scale and replay \adae traces offline.
Figure~\ref{fig:draw-camtraparchitecture} illustrates our framework.
Abstract representations of the edge resources in terms of 
compute, network latency, memory capacity per node, hosted \adae workflows, and workflow partitioning and placement strategies are provided as input.
Our framework automatically partitions representative workflows using distributed, model-parallel slicing techniques~\cite{hsu2019couper,li2023alpaserve}.
The workflows are distributed on representative hardware, which receives visual data from a workload generator that replays and 
scales traces from prior studies.  
Our framework runs \adae workflows on actual hardware
(mainly because \adae studies use affordable and accessible devices),
but it could be adapted to use virtual resources~\cite{Fogify}.
%
Note that our framework is designed only to facilitate the 
study of \adae under realistic conditions. 
State-of-the-art edge simulation platforms provide enhanced 
features and faster setup and execution~\cite{Fogify}.
\subsection{Hardware and network infrastructure}
We utilize six Raspberry Pi 4B units as our edge devices, each equipped with a quad-core ARM Cortex-A72 CPU (1.5GHz), 8GB LPDDR4-3200 SDRAM, and an integrated GPU for basic acceleration tasks. 
ADAE studies frequently use Raspberry Pi units~\cite{Whytock_Suijten_van, Jolles_2021, Fergus_Chalmers_Longmore_Wich_Warmenhove_Swart_Ngongwane_Burger_Ledgard_Meijaard_2023}. 
We seek to mirror these setups. We implemented a custom networking framework to emulate the networking environment encountered in field studies.
We utilize the Linux Traffic Control (tc) utility to emulate various network conditions, including bandwidth limitations, latency, and packet loss characteristics of cellular and satellite links in remote areas. 
This allows us to simulate various real-world networking scenarios, from high-bandwidth, low-latency connections to unreliable, high-latency satellite links.

\subsection{Realistic workload generation}
We used time-stamped traces for \adae datasets collected for 
prior studies.
These studies include camera traps for species distributions \cite{Vélez_McShea_2023} and drones for monitoring wildlife behavior \cite{kholiavchenko2024kabr}.
Timestamps provide the arrival rate for data 
from camera traps and drones that trigger \adae workflow execution.
We are interested in how our system performs as these workloads scale; however, naively increasing the arrival rates can significantly alter the workload's critical characteristics. 
We use correlated or independent scaling, described in Section \ref{sect:scaling}, depending on how the specific workload is expected to scale in a real-life deployment. 
The independent scaling approach involves randomly interleaving bursts and reduces burstiness by increasing the number of independent bursts.
The correlated scaling approach maintains burstiness by appending requests during bursts, effectively increasing the send rate without extending the timeframe.

\subsection{Intelligent model splitting and placement strategies}
\adae workloads are bursty, and previous studies have shown such workloads benefit from pipeline parallelism, although these studies have been restricted to homogeneous hardware \cite{AlpaServe}. 
\adae studies primarily rely on heterogeneous hardware; thus, we focused on load-balancing placement techniques that enable pipeline parallelism on heterogeneous hardware. 
We focus on basic load balancing techniques as a first step to demonstrate that edge computing techniques can be applied to allow for the deployment of \adae studies. 

Edge AI systems designed for \adae study deployment must possess four characteristics to be effective: 
\begin{enumerate}
    \item Ability to exploit bursty workloads
    \item Designed for remote regions with limited compute and memory resources. 
    \item Support latency-sensitive AI computer vision tasks
    \item Run efficiently on diverse edge hardware
\end{enumerate}
We examine four model splitting and placement strategies that accomplish the abovementioned goals, summarized in Table \ref{tab:modelsplitting}. 

\begin{table*}
    \centering
    \begin{tabular}{p{2.5cm}p{3cm}p{4cm}p{3cm}p{3cm}}
    \toprule
    \textbf{Placement Strategy} & \textbf{Description} & \textbf{Procedure} & \textbf{Advantages} & \textbf{Limitations} \\
    \midrule
        Single Node & Baseline approach & One device handles all inference & Simple to implement & Under-utilization of resources, inability to handle large models, lack of adaptability \\
         Round Robin & Assigns whole models to nodes in a round-robin fashion & Each node is assigned one entire model, cycling through the available nodes until all models are placed & Simple to implement and can provide basic load distribution & Does not account for heterogeneous device capabilities or varying model sizes \\
         Equal Utilization & Aims to balance node utilization through intelligently placing splits & Models are split into segments, and these segments are distributed across nodes in a greedy fashion to achieve even utilization & Can lead to reduced queue times and more efficient resource use & Not always optimal if network latency is high or if utilization is low   \\
         Max Distribution & Splits each model across as many nodes as possible & Each model is split into segments proportional to the computational power of the available nodes & Can improve performance when
        load is concentrated on one neural network at a time & Does not scale well with utilization or network latency \\
    \bottomrule \\
    \end{tabular}
 \caption{Edge AI model placement strategies}
    \label{tab:modelsplitting}
\end{table*}

The deployment of AI models in edge computing for \adae studies often follows a naive approach, which we consider our baseline.
In the naive approach, the entire model is placed on individual edge devices without consideration for the specific capabilities of each device or the nature of the workload~\cite{Whytock_Suijten_van, Zualkernan_Dhou_Judas_Sajun_Gomez_Hussain_2022}.
This naive approach has several drawbacks: underutilization of resources, inability to handle large models, and lack of adaptability.
Some devices may be overwhelmed while others remain underutilized, leading to inefficient use of the overall system resources. 
This can result in bottlenecks at specific nodes while others sit idle, reducing the system's overall efficiency.
Edge devices with limited memory may not be able to accommodate larger, more complex models that could provide higher accuracy.
This constraint can force researchers to use simpler, less accurate models, potentially compromising the quality of their ecological insights.
Finally, this static placement cannot adjust to changing workload patterns or network conditions. 
The inability to adapt to dynamic conditions can lead to sub-optimal performance, especially in long-term deployments where environmental and animal behavior patterns may change over time.

To address the limitations of the baseline approach and better meet SLOs, we propose exploring the following strategies: \textit{Naive Round Robin, Utilization-Balanced Model Splitting}, and \textit{Proportional Model Splitting}, shown in Table~\ref{tab:modelsplitting}. 
The naive round-robin approach does not split models but instead assigns whole models to available nodes.
This approach is simple to implement and can provide load distribution. However, it does not account for heterogeneous device capabilities or varying model sizes.
The utilization-balanced model splitting approach balances node utilization through intelligent model layer splitting, using a bin-packing algorithm~\cite{hsu2019couper} to determine optimal splitting and placement.
To implement proportional model splitting, each model is split into segments proportional to the computational power of the available nodes.
This approach may not achieve perfectly balanced node utilization but can provide better throughput for collocated workloads that do not receive traffic simultaneously.

\begin{figure*}[t]
    \centering
    \includegraphics[scale=0.4]{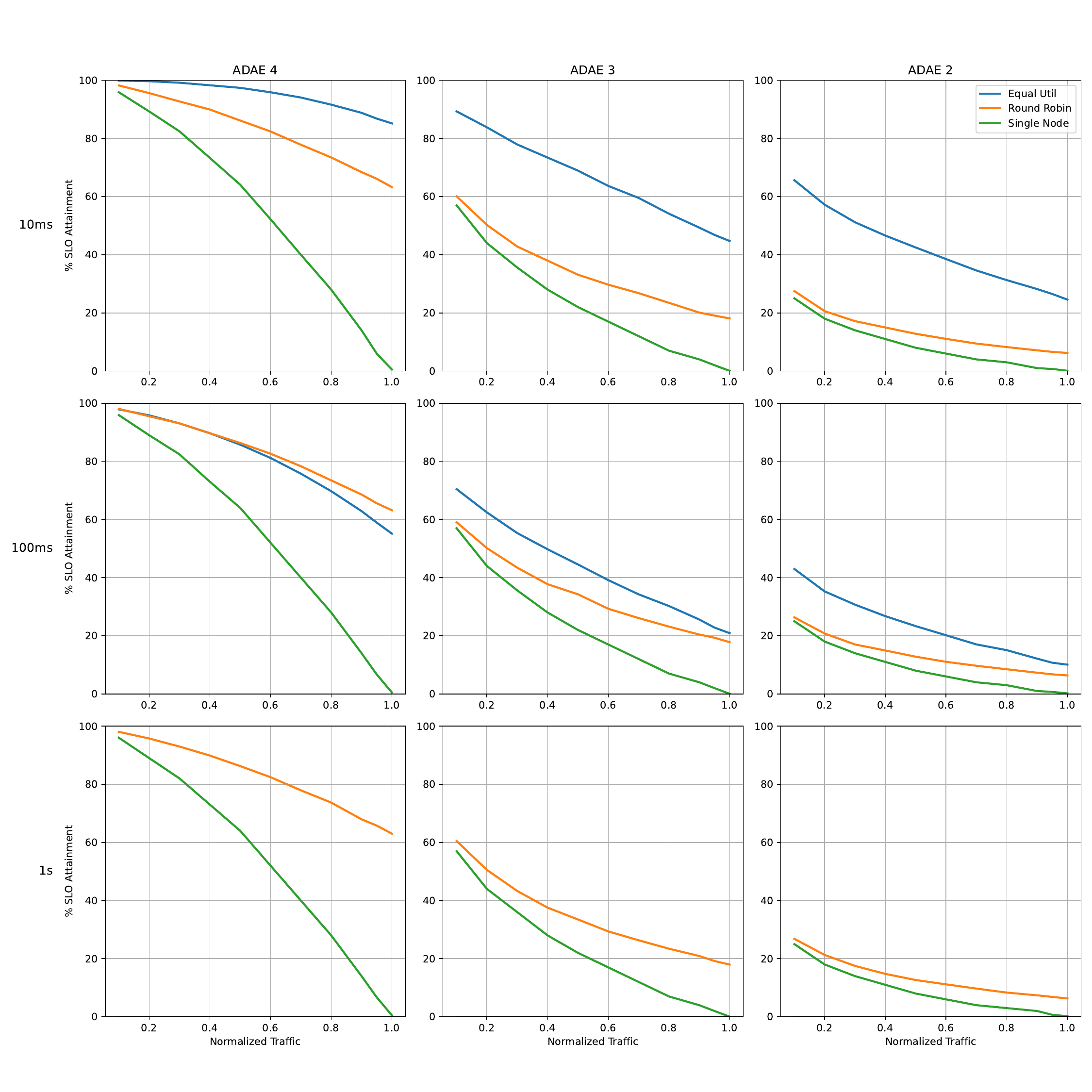}
    \caption{Performance comparison of different placement strategies (Equal Utilization, Round Robin, Single Node) under various ADAE workloads (each column) and network latencies (each row). The tests were conducted using a 2-node cluster. The x-axis (normalized traffic) is the utilization level of a single node, i.e. the arrival rate of a single node. }
    \label{fig:combined-results}
\end{figure*}

\begin{figure*}[t]
    \centering
    \includegraphics[scale=0.4]{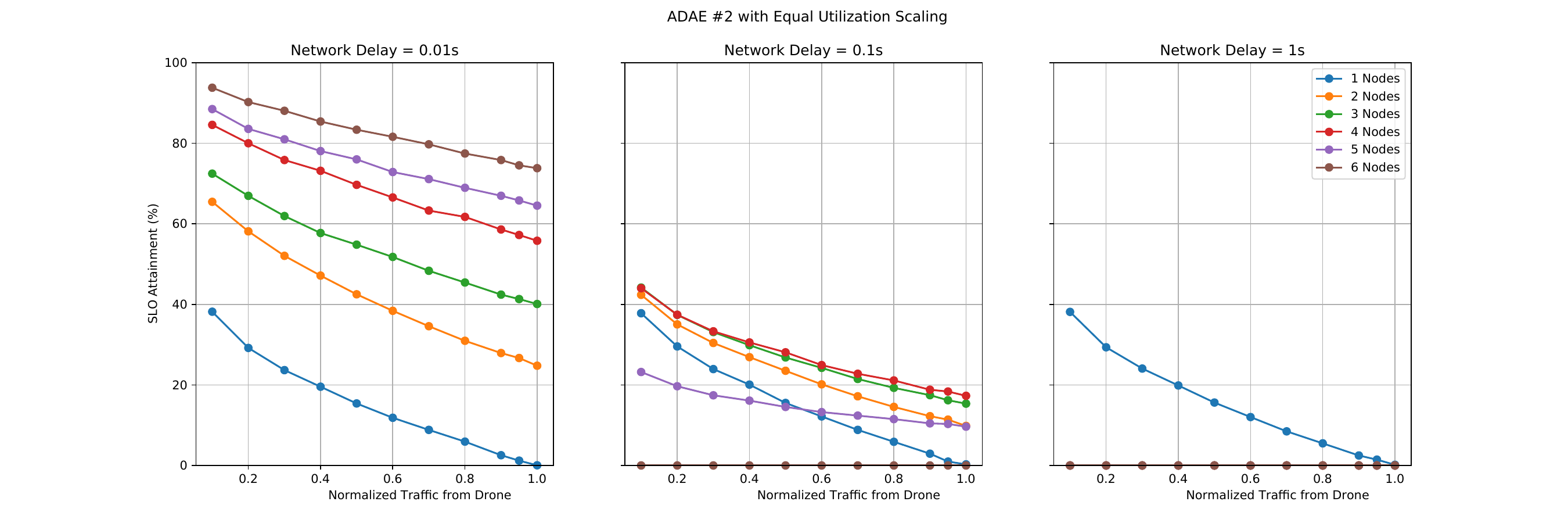}
    \caption{Scaling performance of the equal utilization placement strategy under different network latencies (0.01s, 0.1s, 1s) for the ADAE 2 workload. The ADAE 2 workload was selected due to its relatively lower SLO attainment on a 2-node setup. Each line represents a different number of compute nodes (ranging from 1 to 6).}
    \label{fig:EqualUtilScalingNodes}
\end{figure*}

\subsection{Representative computer vision models}
Our experiments are conducted using the YOLOv5 \cite{yolov8_ultralytics} suite of models.
We evaluated YOLO since it is currently one of the most popular and widely used models for \adae computer vision tasks including detection, classification, and behavior identification. 
The YOLOv5 family includes models of varying sizes and complexities, allowing us to evaluate our strategies across a spectrum of computational demands. 
This choice reflects the common use of YOLO-based models in recent \adae studies due to their efficiency and accuracy in real-time object detection, localization, and classification tasks \cite{kline_acsos, Xie_Jiang_Bao_Zhai_Zhao_Zhou_Jiang_2023}.

\subsection{Experimental procedure}
Our experiment procedure has four steps: establish a baseline, evaluate the strategy, compare to the baseline, and simulate network conditions. 
We establish baseline performance metrics for each model and workload combination using standard, non-distributed inference. 
This provides a point of comparison to quantify the improvements achieved by our proposed strategies.
We systematically evaluate our three distributed inference strategies: \textit{Naive Round Robin, Utilization-Balanced Splitting}, and \textit{Proportional Splitting}. 
Each strategy is tested across workload scenarios and network conditions to assess its robustness and adaptability.
We implement and compare the three placement strategies across various workload scenarios, focusing on their ability to meet the defined \slo. 
This comparison helps identify the strengths and weaknesses of each approach under different operating conditions.
We simulate different inter-node and intra-node network conditions to evaluate the strategies' performance under various connectivity scenarios typical in animal ecology field studies. 
This includes testing under ideal conditions and challenging scenarios with high latency and low bandwidth.

\subsection{Performance metrics}
We collect comprehensive metrics to evaluate system performance, focused on latency and resource utilization.
For latency, we capture end-to-end processing time for individual inference requests, including network transmission delays. 
This metric is crucial for assessing the system's ability to provide real-time adaptations, which is vital for \adae.
We measure the utilization of CPU, GPU, and memory across all devices in the cluster. 
These metrics assess the efficiency of our placement strategies in balancing load across heterogeneous resources.
We evaluate how well each strategy achieves \slo attainment, which captures the percentage of inference requests meeting predefined latency thresholds under each \adae from Table~\ref{table:ae_slos}.
For resource utilization balance, variance in CPU, GPU, and memory utilization across nodes for each strategy. 
A low variance indicates more balanced resource utilization, which can lead to better overall system efficiency and reduced bottlenecks.
We will assess how each strategy performs under varying network conditions, measuring the degradation in performance as network quality decreases. 
This analysis will help identify which strategies are most robust to the challenging and variable network environments often encountered in remote field studies.

\subsection{Characterizing performance of \adae studies on representative edge hardware}
Figure~\ref{fig:combined-results} shows the effects of burstiness, network latency, and placement strategy on \slo attainment as the arrival rate of data increases (i.e., normalized traffic).
We tested two co-located workloads for all experiments.
The first workload comprising 30\% of the aggregate traffic is \adae 1.  The other co-located workload is shown in Figure~\ref{fig:combined-results}. 
The bottom row shows performance under slow network connectivity at the edge with a 1-second round trip time.
As expected, the distributed inference is ineffective in this context, and the worst-performing strategy is the equal-utilization policy.  
In contrast, the top row shows equal utilization consistently outperforms all other policies under fast-edge networks. 
 \adae 4 is the most bursty and \adae 2 is the least.
Looking across burstiness in the columns, we observe that burstier workloads magnify the performance gains achieved by the placement strategies. 

Figure~\ref{fig:EqualUtilScalingNodes} depicts the effect of increasing edge resources as the arrival rate increases. 
As expected under slow network latency, the equal utilization placement 
strategy performed poorly.  
However, under low-latency network configurations, this approach
achieves near-linear scaling.  
Finally, we also tested how equal utilization placements compare to maximum distribution when animals show avoidant behavior, increasing burstiness. 
We observed a slight but consistent improvement of roughly 5\% for maximum distribution at scale.

\section{Performance Modeling for \adae Studies }
\label{sect:results}
We tested four models for estimating the performance gains for a given edge AI system: random forest, XD Gradient, M/D/1 queuing model, and a hybrid random forest, M/D/1 queuing model, shown in Table \ref{tab:accuracy}.
 XD Gradient boost has performed well in previous studies in predicting system performance with few data points. 
For our study, however, XD Gradient only produced a 35\% accuracy in estimating performance gains.
We also tested a regular M/D/1 queue, assuming Poisson arrival rates, which also produced a 30\% error.
The random forest takes the utility level, $\lambda$, $\mu$, and expected output latency as inputs, which generated a 24\% error.
The optimal model for predicting performance gains was a hybrid random forest, M/D/1 queue approach, which produced a 19.6\% error.
This model predicts expected random forest performance gains and fine-tunes the results with the M/D/1 queue to predict the number of anticipated bursts to overlap for a given workload.

\begin{table}[h]
    \centering
    \begin{tabular}{ll}
    \toprule
         \textbf{Model}& \textbf{Error Rate}\\
         \midrule
         XD Gradient& 35 \%\\
         M/D/1 Queue& 30 \%\\
         Random Forest& 24 \%\\
         Random Forest M/D/1 Hybrid& 19 \% \\
         \bottomrule \\
    \end{tabular}
    \caption{Distributed Edge Provisioning Techniques}
    \label{tab:accuracy}
\end{table}

We expected to see a reduction in error rates with additional data points.
The advantage of this hybrid approach over previous works, such as AlpaServe \cite{AlpaServe}, is that this method allows practitioners to create a scheduling system without requiring historical data, using traces from a single deployed node, optimized for their specific Edge AI system and \adae.

\section{Related Work}
\label{sect:relatedwork}

Edge computing and artificial intelligence are increasingly being applied to ADAE studies, enabling new data collection and analysis approaches. 
This convergence of technologies, often referred to as Edge AI, has the potential to enable sophisticated processing of data gathered from remote sensing devices such as camera traps and drones. 
Edge AI systems perform computations at the edge near the source of the data, as opposed to sending data to a centralized cloud server \cite{Singh_Gill_2023}.
Edge AI requires massive amounts of data and computing capacity. Still, recent advancements in sensors, hardware, and communication technology like 5G and 6G networks have made this possible on the edge in remote regions \cite{Singh_Gill_2023, Jolles_2021}.

Edge AI is enabled by distributed computing paradigms that allocate tasks across a network of devices.
Recent studies have demonstrated how model splitting and co-location can be applied to edge computing paradigms to improve system performance~\cite{PipeEdge, ramprasad2022shepherd, Hao_Zhang_2021}.
Model splitting and co-location can reduce latency and utilize system compute more efficiently, i.e., increase the frequency at which the system achieves its \slo~\cite{AlpaServe, PipeEdge}. 
Implementing model splitting and co-location has effectively reduced latency for bursty workloads, although this study focused on homogeneous hardware ~\cite{AlpaServe}.
Heterogeneous hardware and network conditions are considered in~\cite{Hou_Guan_Han_Zhang_2022}, which proposes a deep reinforcement learning approach to speed up convolutional neural network inference on distributed edge devices.
Edge AI system performance can be optimized through model architecture for distributed inference~\cite{Giovannesi_Proietti_Mattia_Beraldi_2024, Eccles_Wong_Varghese_2024, Shuvo_Islam_Cheng_Morshed_2023}.

Optimizing camera placement to improve the performance of computer vision pipelines has been investigated, namely for traffic cameras \cite{Wong_Ramanujam_Balakrishnan_Netravali}. 
Sensor position and orientation dictate the data that cameras can capture, which in turn dictates the accuracy of real-time image analytics.
Edge AI enables sensors to be continuously adjusted in real time to maximize workload accuracy under resource constraints.
Traffic tasks are similar to animal ecology computer vision tasks. They include detecting objects of interest, counting the number of objects of interest, detection with bounding boxes, and aggregate counting of unique objects of interest \cite{Wong_Ramanujam_Balakrishnan_Netravali}.
However, this traffic camera study is restricted to a single camera and does not consider request arrival rates.
ADAE studies must consider the network of sensors, including drone and camera traps, as well the characteristically bursty arrival rates when designing and implementing runtime adaptations.

As the volume and complexity of ecological data continue to grow, there is an increasing need for efficient computing approaches that can handle the unique challenges posed by wildlife monitoring in remote environments.
Recent studies have investigated on-device processing for a more immediate analysis of ecological data.
Mobile computing devices, such as laptops, tablets, or custom-built portable units, may also augment in-situ processing capabilities. 
Such devices could serve as intermediate processing nodes, bridging the gap between data collection points and cloud infrastructure \cite{Gholami_Baras_2021}. However, more research is needed to establish their effectiveness in field conditions \cite{Jolles_2021, Whytock_Suijten_van}.

Ongoing improvements in the performance of edge processors may enable the deployment of more sophisticated AI models on smart camera traps and drones.
Smart camera traps, equipped with on-board computers, can perform initial data processing and filtering, reducing the volume of data that needs to be transmitted or stored for later analysis\cite{Ahumada_Fegraus_Birch_2020, Zualkernan_Dhou_Judas_Sajun_Gomez_Hussain_2022, Whytock_Suijten_van, Dertien_Negi_Dinerstein_Krishnamurthy_Negi_Gopal_Gulick_Pathak_Kapoor_Yadav_2023, Ma_Dong_Xia_Xu_Xu_Chen_2024, Fergus_Chalmers_Longmore_Wich_Warmenhove_Swart_Ngongwane_Burger_Ledgard_Meijaard_2023}.
Drones equipped with on-board GPU are increasingly available, enabling real-time, on-board processing for autonomous navigation policies~\cite{kline_acsos, Luo_Zhang_Shao_2024, Andrew_Greatwood_Burghardt_2020, Andrew_Greatwood_Burghardt_2019}.
Ecologists have raised concerns about the potential risks of disturbance of wildlife caused by drones \cite{schaul2015prioritized}, which could be reduced by edge-enabled autonomous navigation equipped with safeguards.

The AI and ecology communities have a history of collaboration, applying state-of-the-art computer vision techniques to uncover ecological insights.
The CV4Animals: Computer Vision for Animal Behavior workshop, held annually at The Conference on Computer Vision and Pattern Recognition (CVPR), published 40 works this past year alone and featured 18 previously published works on computer-vision-based animal behavioral analysis.
As computer vision models grow in size and complexity, we expect models developed for ADAE applications to follow this trend.
CNN-based models such as YOLO \cite{yolo_2023} remain popular for detection, localization, and classification tasks for ADAE studies~\cite{kholiavchenko2024kabr, kline_acsos, Xie_Jiang_Bao_Zhai_Zhao_Zhou_Jiang_2023}, and YOLO-based models have been tuned to boost performance on \adae aerial imagery \cite{Mou_Liu_Zhu_Cui_2023}.
Recently, vision transformer (ViT) models have proven to perform well, particularly for multi-modal foundation models, such as the species classification models BioClip~\cite{stevens2024bioclip} and Arboretum~\cite{yang2024arboretumlargemultimodaldataset}, both based on the OpenCLIP~\cite{ilharco_gabriel_2021_5143773} ViT architecture. 
Increasingly sophisticated models have been developed for more specialized ADAE studies, including inferring animal behavior from video \cite{Brookes_Mirmehdi_Kühl_Burghardt_2023} and 3D pose estimation of wildlife from drone footage \cite{Shukla_Morelli_Remondino_Micheli_Tuia_Risse_2024}. 

\section{Conclusion and Future Work}
\label{sect:conclusion}

\adae studies have the potential to revolutionize animal ecology and biodiversity studies.
Unlike traditional ecological studies in the field, \adae studies leverage edge computing resources to control smart camera traps and autonomous drones, filtering images and adjusting the viewing angles at runtime to improve data quality.
Runtime adaptations improve data quality, allowing ecologists to derive insights quickly from their data. 
They also reduce the time spent parsing data irrelevant to study objectives.
Data captured after runtime adaptations can differentiate between datasets that yield insights and inconclusive studies.
For these reasons, \adae studies are a growing edge workload.
%

%
An essential contribution of this work is discovering commonalities of \adae traces that enabled rigorous analysis.
Using timestamped traces from prior studies, we observed that (1) the workflows are characterized by interdependent, complex  computer vision tasks that transform harvested visual data into ecological datasets;
(2) \slo can be repurposed to describe the strict latency  demands required for runtime adaptation; and (3) animal dynamics partially explain the 
bursty workloads observed across many studies.

We replayed ADAE traces offline on representative hardware to understand interactions with edge hardware.
We found that workflow partitioning schemes have a complex effect on SLO attainment, especially at scale.   
We also found that performance modeling approaches using queuing theory and machine learning provide a good starting point to predict SLO attainment.  

AI models will likely increase in complexity following current trends. 
However, \adae studies are still in the very early stages of adoption.
We anticipate simple models, such as YOLO, will be the first to be implemented in real time to inform system adaptations. 
Thus, we first focused on profiling the implementation of ADAE studies using YOLO.
We will expand our approach to implement more complex AI models and workflows in the future.
We plan to explore more sophisticated load-balancing approaches tailored to the specific application and available edge devices.
We hope others will be interested in investigating this as well.

We encourage others to leverage our findings to propose innovative edge systems for sophisticated \adae to further our ability to understand and protect our planet's biodiversity.
Numerous interdisciplinary innovations have been in computer vision and ecology, but these sophisticated AI models require edge computing innovation to be successfully deployed.
\adae studies offer transformative potential for animal ecology by using edge computing to control smart camera traps and drones, enhancing biodiversity research through advanced edge systems.

\section{Acknowledgments}
The ICICLE project is funded by the National Science Foundation (NSF) under grant number OAC-2112606. The NSF supports the Imageomics Institute under Award No. 2118240.

\bibliographystyle{plain}  
\bibliography{ref}

\end{document}

%% file: figures/table.tex
\begin{sidewaystable}

\footnotesize
\begin{tabular}{p{0.1cm}p{1.5cm}p{1.5cm}p{1cm}p{0.5cm}p{3cm}p{3.5cm}p{0.5cm}p{0.5cm}}
 \toprule
 & \multicolumn{1}{c}{} & \multicolumn{1}{c}{} & \multicolumn{1}{c}{} & \multicolumn{1}{c}{} & \multicolumn{1}{c}{} & \multicolumn{1}{c}{} & \multicolumn{2}{c}{\textbf{SLO}} \\
 & \multicolumn{1}{c}{} & \multicolumn{1}{c}{} & \multicolumn{1}{c}{} & \multicolumn{1}{c}{} & \multicolumn{1}{c}{} & \multicolumn{1}{c}{} & \multicolumn{1}{c}{\textbf{Latency}} & \multicolumn{1}{c}{} \\
\multirow{-3}{*}{\textbf{ID}} & \multicolumn{1}{c}{\multirow{-3}{*}{\textbf{Location}}} & \multicolumn{1}{c}{\multirow{-3}{*}{\textbf{Hardware}}} & \multicolumn{1}{c}{\multirow{-3}{*}{\textbf{Species}}} & \multicolumn{1}{c}{\multirow{-3}{*}{\textbf{Data}}} & \multicolumn{1}{c}{\multirow{-3}{*}{\textbf{AI Computer Vision Tasks}}} 
& \multicolumn{1}{c}{\multirow{-3}{*}{\textbf{\adae Result}}} 
& \multicolumn{1}{c}{sec/frame} 
& \multicolumn{1}{c}{\multirow{-2}{*}{\textbf{Requests Met}}} \\
\midrule \\
1 & Yellowstone Park, USA & Single fixed-wing drone & Bison & Photo & Detect, localize & Count of calves in herd & 0.4 & 50\% \\
2 & Zimbabwe & Single-fixed wing drone & Multiple species & Photo & Extract frames, detect, localize, classify  & Detect endangered species & 1.0 & 99\% \\ 
3 & Kenya & Single quadcopter & Giraffe & Photo & Detect, localize, classify, track & Count by habitat type & 1.0 & 80\% \\
4 & Kenya & Quadcopter swarm & Zebra & Video & Extract frames, detect, localize, track  & Behavioral time budgets & 1.0 & 80\% \\ 
5 & Conservation Center, USA & Single smart camera trap & African Wild Dogs & Video & Extract frames, detect, localize, track & Behavior by time of day & 0.03 & 95\% \\
6 & Columbia & Single smart camera trap & Multiple Species & Photo & Detect, localize, classify & Species distribution and population estimates & 180 & 99\% \\
7 & Columbia & Camera trap network & Multiple Species & Video & Detect, localize, classify, track & Behavior \& individual identification & 0.03 & 95\%\\
\bottomrule \\
\end{tabular}
\caption{AI-driven animal ecology (\adae) studies service-level objectives (SLOs)}
\label{table:ae_slos}
\end{sidewaystable}